\documentclass[letterpaper,twocolumn,prb,aps,showpacs,floatfix, 
superscriptaddress]{revtex4-2}
\usepackage{bbm}
\usepackage{soul}
\usepackage{bm}
\usepackage[usenames]{color}
\usepackage{multirow}
\usepackage{amssymb}
\usepackage{amsbsy}
\usepackage{mathtools}
\usepackage{amsmath}
\usepackage{stmaryrd}
\usepackage{graphicx}
\usepackage{epsfig}
\usepackage{placeins}
\usepackage{bbold}
\usepackage{braket}
\usepackage{blindtext}
\usepackage[colorlinks,linkcolor=blue,citecolor=cyan,urlcolor=blue]{hyperref}
\usepackage{scalerel}
\usepackage{tikz}
\usetikzlibrary{calc}
\usetikzlibrary{patterns}
\usetikzlibrary{svg.path}
\definecolor{orcidlogocol}{HTML}{A6CE39}
\tikzset{
  orcidlogo/.pic={
    \fill[orcidlogocol] 
svg{M256,128c0,70.7-57.3,128-128,128C57.3,256,0,198.7,0,128C0,57.3,57.3,0,128,
0C198.7,0,256,57.3,256,128z};
    \fill[white] svg{M86.3,186.2H70.9V79.1h15.4v48.4V186.2z}
                 
svg{M108.9,79.1h41.6c39.6,0,57,28.3,57,53.6c0,27.5-21.5,53.6-56.8,
53.6h-41.8V79.1z 
M124.3,172.4h24.5c34.9,0,42.9-26.5,
42.9-39.7c0-21.5-13.7-39.7-43.7-39.7h-23.7V172.4z}
                 
svg{M88.7,56.8c0,5.5-4.5,10.1-10.1,10.1c-5.6,0-10.1-4.6-10.1-10.1c0-5.6,4.5-10.1
,10.1-10.1C84.2,46.7,88.7,51.3,88.7,56.8z};
  }
}

\newcommand\orcid[1]{\!%
  \href{https://orcid.org/#1}{%
    \mbox{%
      \scaleto{%
        \begin{tikzpicture}[yscale=-1,transform shape]
          \pic{orcidlogo};
        \end{tikzpicture}
      }{8pt}%
    }%
  }%
}

\begin{document}
\title{Fate of diffusion under integrability breaking of classical integrable magnets}

\author{Jiaozi Wang~\orcid{0000-0001-6308-1950}}
\email{jiaowang@uos.de}
\affiliation{Institute of Fundamental Physics and Quantum Technology, and
School of Physical Science and Technology, Ningbo University, Ningbo, Zhejiang 315211, China}
\affiliation{Department of Mathematics/Computer Science/Physics, University of Osnabr\"uck, D-49076 
Osnabr\"uck, Germany}

\author{Sourav Nandy~\orcid{0000-0002-0407-3157}}
\affiliation{Max Planck Institute for the Physics of Complex Systems, D-01187
Dresden, Germany}
\author{Markus Kraft~\orcid{0009-0008-4711-5549}}
\affiliation{Department of Mathematics/Computer Science/Physics, University of Osnabr\"uck, D-49076 
Osnabr\"uck, Germany}

\author{Toma\v{z} Prosen~\orcid{0000-0001-9979-6253}}
\affiliation{Faculty of Mathematics and Physics, University of Ljubljana, Jadranska 19, SI-1000 Ljubljana, Slovenia}
\affiliation{Institute of Mathematics, Physics and Mechanics, Jadranska 19, SI-1000 Ljubljana, Slovenia}

\author{Robin Steinigeweg~\orcid{0000-0003-0608-0884}}
\email{rsteinig@uos.de}
\affiliation{Department of Mathematics/Computer Science/Physics, University of Osnabr\"uck, D-49076 
Osnabr\"uck, Germany}

\date{\today}


\begin{abstract}
Diffusive transport is a ubiquitous phenomenon, yet the microscopic origin of diffusion in interacting physical systems remains a challenging question, irrespective of whether quantum effects are dominant or not. In this work, we study infinite temperature  spin diffusion in a classical integrable, space-time discrete version of anisotropic Landau-Lifshitz magnet in the easy-axis regime, subjected to integrability-breaking perturbations. Our numerical results based on large-scale simulations reveal i) a sharp change in the spin diffusion constant as a function of perturbation strength in the thermodynamic limit and ii) a crossover from non-Gaussian to Gaussian statistics of magnetization transfer reflected in higher order cumulants under integrability breaking. Both our observations hint to the presence of non-trivial diffusion mechanism inherent to integrable systems.

\end{abstract}
\maketitle


\underline{\it Introduction:}\  Non-equilibrium processes are ubiquitous in nature and form a central theme of modern physics research \cite{mazenko2008nonequilibrium, 2013nmtq.book.....S,D'Alessio03052016-NE,eisert2015quantum-NE,RevModPhys.80.885-NE,RevModPhys.83.863-NE,RevModPhys.91.021001-NE, altman2015nonequilibriumquantumdynamics}. A fundamentally important example is  transport phenomenon \cite{mazenko2008nonequilibrium, Bertini_RevModPhys}, with diffusion being one of the most extensively studied processes since Fourier's seminal formulation of its continuum description \cite{Fourier}. Despite centuries of intense investigation, one of the central challenges of statistical mechanics remains understanding how irreversible phenomena such as diffusion emerge from fundamentally time-reversible microscopic laws \cite{landau2013statistical}. While addressing such question typically requires suitable approximations, integrable systems provide a  platform to compute physical quantities in an exact or at least in a controlled manner \cite{Takahashi:1999bgb, Eckle}.

While diffusion is expected to  characterize high-temperature transport in generic interacting systems, its emergence in integrable models is highly non-trivial \cite{Jacopo_diffusion_PRL2018, Sarang_PRB2018_diffusion,Bertini_RevModPhys}. In the paradigmatic spin-1/2 XXZ chain, decades of extensive studies based upon Bethe ansatz and formalism of quantum integrability \cite{Tomaz_PRL_2013, Tomaz_PRL_2011}, generalized hydrodynamics \cite{Jacopo_diffusion_PRL2018, GHD_diffusion_Jacompo_1, GHD_a_perspective, Bulchandani_2021}, and advanced numerical methods (e.g., dynamical quantum typicality \cite{Robin_DQPT_2014, Robin_diffusion_2015, Robin_diffusion_2017,PhysRevLett.102.110403-DQT,doi:10.7566/JPSJ.90.012001-DQT,HeitmannRichterSchubertSteinigeweg-DQT} and approaches based on tensor network ansatz \cite{Marko_diffusion_prl_2011, Christoff_2014_diffusion, Karasch_2012_PRL}) indicate normal spin diffusion in the easy-axis regime at zero magnetization. However, more refined analyses using full counting statistics (FCS) reveal that this regime also exhibits signatures of anomalous spin diffusion, as higher-order moments of the magnetization transfer distribution display non-Gausssian features \cite{Marko_2014_Cumulants, nonGaussian-XXZ-25, PhysRevLett.128.090604_Model,PhysRevLett.132.017101_Model}.

Given the manifold nature of diffusion in integrable systems and the fine tuned character of integrability itself, it is crucial to understand how diffusion behaves under integrability breaking perturbations. Such perturbations render the system generic, raising key questions: To what extent do anomalous signatures of diffusion, for example those evident in full counting statistics, survive weak breaking of integrability? How does the diffusion constant scale with perturbation strength? In particular, can this scaling reveal a crossover from anomalous diffusion characteristic of integrable systems to normal diffusion in generic many body systems? These are the questions we address in this work.

A quantitative answer of the above questions, however, demands a careful finite-size scaling analysis, which is a formidable task in quantum many-body systems due to the curse of Hilbert-space dimensionality. To overcome this challenge, we turn to classical integrable magnets, which provide a controlled and computationally tractable framework where integrability and its weak breaking can be explored over large system sizes. In this work, we focus on the anisotropic Landau-Lifshitz magnet (LLM), a paradigmatic classical integrable system that exhibits the same dynamical phases as the spin-1/2 XXZ chain. In addition, the aforementioned model is known to be the effective evolution law for the semi-classical eigenstates in the quantum spin chain \cite{Enej_QC_2021SciPost, Enej_QC_2020PRL}.

Building on the aforementioned framework, we investigate magnetization transport in the  anisotropic LLM subject to (weak) integrability-breaking perturbations. 
Using large-scale numerical simulations, we determine the scaling behavior of diffusion constant with system size and perturbation strength.
Our result reveals a discontinuous change of the diffusion constant when the system departs from the integrable point.
Furthermore, analysis of the higher order cumulant of the magnetization current distributions shows the restoration of Gaussian statistics (and thereby of normal diffusion) upon integrability breaking. Notably, these results are consistent with previous observations made in quantum spin-1/2 XXZ chain \cite{DeNArdis_PNAS, Nandy_PRB2022, nonGaussian-XXZ-25}, while the classical setting allows for highly controlled and rigorous numerical verification, which renders the behavior clearer and more convincing. In addition, we found evidence of universal finite time and system size scaling of diffusion constant with perturbation strength. Our findings may provide an non-trivial example of classical-quantum correspondence in transport phenomena\cite{PhysRevB.99.140301-qc,PhysRevB.104.054415-qc,PhysRevB.111.064420-qc,PhysRevResearch.2.013130-qc}.

\underline{\it Model:}\  
We consider the anisotropic LLM, which in the continuous space-time is described as the equation of motion for spin field $\boldsymbol{S}_x \equiv (S^1_x,S^2_x,S^3_x) \in {\cal S}^2$ as
\begin{equation}\label{eq-LLC}
\partial_{t}\boldsymbol{S}=\boldsymbol{S}\times\partial_{x}^{2}\boldsymbol{S}+\boldsymbol{S}\times\boldsymbol{JS},
\end{equation}
where the anisotropy tensor $\boldsymbol{J}=\text{diag}(0,0,\delta)$ and $\delta \in \mathbb{R}$. Eq.~\eqref{eq-LLC} is an integrable partial differential equation (PDE) \cite{TAKHTAJAN1977235} and the third component of the total spin
$Q = \int dx\, S^3(x,t)$ is one of the infinitely many conserved charges that the PDE hosts. The spin density $S^3(x,t)$ satisfies the continuity equation
$\partial_t S^3(x,t) + \partial_x j(x, t) = 0,$ where $j(x, t)$ denotes the current density at the position $x$, time $t$ and the expression of the current is given in the End Matter (EM).
In practice, our numerical simulation is performed on discrete space-time lattice version of Eq. \eqref{eq-LLC}. A naive discretization destroys integrability and to preserve the same, we employ a special two-parameter (denoted by $\rho$ and $\tau$ in the following) symplectic discretization as discussed in \cite{10.21468/SciPostPhys.11.3.051_model}. 
To this end, we introduce an anisotropic stereographic variable
  $\zeta \in \mathbb{C}$
\begin{equation}
\zeta=\sqrt{\frac{\sinh[\rho(1-S^{3})]}{\sinh[\rho(1+S^{3})]}}\frac{S^{-}}{\sqrt{1-(S^{3})^{2}}} ,
\end{equation}
where $S^{-} = S^{1} - iS^{2}$. The local two-body map, e.g., acting on sites $x=1,2$, is given by \cite{krajnik2024integrable_model}
\begin{gather}
\zeta^\prime_1 = \varPhi_{\tau,\rho}(\zeta_{1},\zeta_{2})= \nonumber \\
\frac{\zeta_{1}\sinh(2\rho)+\zeta_{2}|\zeta_{1}|^{2}\sinh[2\rho(1-i\tau)]-\zeta_{2}\sinh(2i\tau\rho)}{\zeta_{2}\overline{\zeta}_{1}\sinh(2\rho)+\sinh[2\rho(1-i\tau)]-|\zeta_{1}|^{2}\sinh(2i\tau\rho)},
\end{gather}
and 
\begin{equation}
    \zeta^\prime_2 = \varPhi_{\tau,\rho}(\zeta_{2},\zeta_{1})\ .
\end{equation}
The global map $\varPsi_{\tau, \rho}$ is then defined as 
\begin{gather}
(\zeta_{1}^{t+1},\zeta_{2}^{t+1},\ldots,\zeta_{L}^{t+1})=\varPsi_{\tau,\rho}(\zeta_{1}^{t},\zeta_{2}^{t},\ldots,\zeta_{L}^{t}),\nonumber \\
\varPsi_{\tau,\rho}=\varPsi_{\tau,\rho}^{\text{odd}}\circ\varPsi_{\tau,\rho}^{\text{even}}=\prod_{x=0}^{L/2-1}\varPhi_{\tau,\rho}^{2x+1}\circ\prod_{x=0}^{L/2-1}\varPhi_{\tau,\rho}^{2x}, \label{eq-oe-map}
\end{gather}
where $\varPhi_{\tau,\rho}^{x}$ indicates local map acting on sites $x$, $x+1$ and
we employ the periodical boundary condition (PBC) $\zeta_{x+L} = \zeta_{x}$ .
The parameter $\rho$ is related to $\delta$ through $\delta = \rho^2$\cite{PhysRevLett.128.090604_Model}.
By tuning the anisotropy parameter $\delta$ (and thus $\rho$ in turn), one accesses three dynamical regimes:
(i) the easy-plane ballistic regime for $\rho = i\gamma,\ \gamma\in[-\pi/2,\pi/2]\  (\delta < 0)$, 
(ii) the isotropic point with super-diffusive transport at $\rho = 0 (\delta = 0)$, and 
(iii) the easy-axis normal diffusive regime for $\rho \in \mathbb{R}_+\  (\delta > 0)$.
Thus, this classical system shares exact correspondence with the dynamical phases of the spin-1/2 XXZ chain \cite{Bulchandani_2021, Gopalakrishnan_2023, Bertini_RevModPhys}.
In this Letter, we focus on the easy-axis regime, and set $\rho = 1$ and $\tau = 1$. 
Within this regime, the system  exhibits diffusive magnetization transport at high temperature \cite{10.21468/SciPostPhys.11.3.051_model}. 


\begin{figure}[t]
	\includegraphics[width=1.0\linewidth]{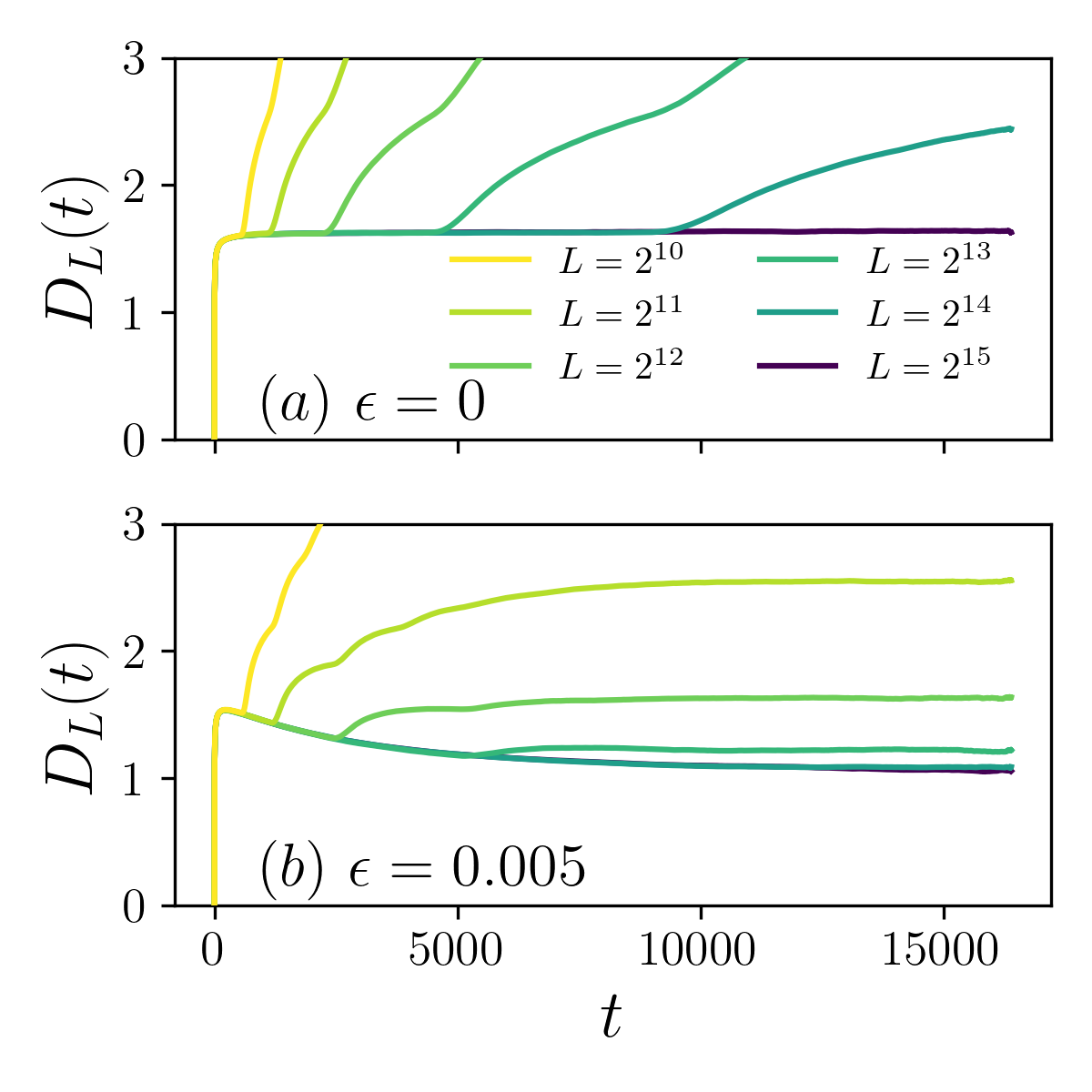}
 \caption{ Time-dependent diffusion constant $D_L(t)$ versus $t$ in model $A$ for different system size $L$ for (a) $\epsilon = 0$ and (b) $\epsilon = 0.005$}\label{Fig1}
\end{figure}

To break integrability of the system, we consider two distinct types of perturbations. The resulting maps read
    \begin{equation}
     \varPsi^{A}_{\tau,\rho}=\varPsi_{\tau+\Delta\tau,\rho}^{\text{odd}}\circ\varPsi_{\tau-\Delta\tau,\rho}^{\text{even}}=\prod_{x=0}^{L/2-1}\varPhi_{\tau+\Delta\tau,\rho}^{2x+1}\circ\prod_{x=0}^{L/2-1}\varPhi_{\tau-\Delta\tau,\rho}^{2x}
    \end{equation}
and 
\begin{equation}
\varPsi^{B}_{\tau,\rho}=R_{\theta}\circ\varPsi_{\tau,\rho}^{\text{odd}}\circ R_{\theta}\circ\varPsi_{\tau,\rho}^{\text{even}},
\end{equation}
where
$R_{\theta}=\prod_{x=1}^{L/2}r_{\theta}^{2x}$,
with $r_{\theta}^{x}$ indicating local rotation map at $x$,
\begin{equation}
        r_{\theta}(\zeta)=e^{-i\theta}\zeta.
\end{equation}
For convenience, we refer to the space-time lattices generated by $\varPsi^{A}_{\tau,\rho}$ and $\varPsi^{B}_{\tau,\rho}$ as Model~A and Model~B, respectively. 
Their corresponding parameters, $\Delta \tau$ and $\theta$, represent the perturbation strengths and are hereafter both denoted by $\epsilon$. 
In what follows, we introduce the quantities of interest that we use to characterize transport features.

\begin{figure}[t]
	\includegraphics[width=1\linewidth]{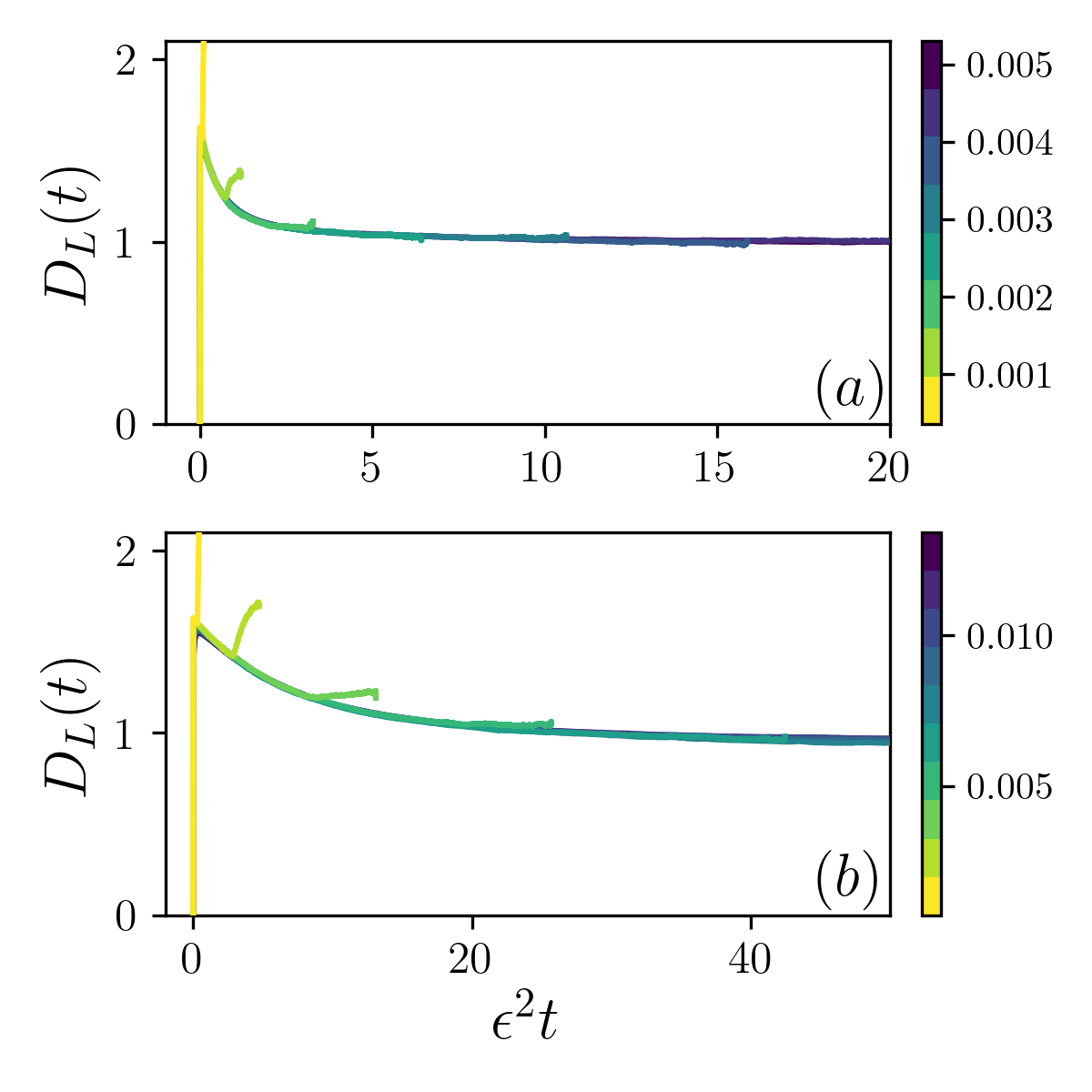}
 \caption{$D_L(t)$ versus $\epsilon^2 t$ for different $\epsilon$ in (a) model $A$ and (b) model $B$.
System size $L = 2^{17}$; $\epsilon \in [0, 0.0053]$ in (a) and $\epsilon \in [0, 0.0141]$ in (b), with color ranging from yellow to blue as $\epsilon$ increases.
 }\label{Fig2}
\end{figure}

{\underline {\it Diffusion Constant:}}\  
Diffusion constant is defined as
\begin{equation}
D=\lim_{t\rightarrow\infty}\lim_{L\rightarrow\infty}D_{L}(t) ,
\end{equation}
where $D_L(t)$ indicates the time dependent diffusion constant at system size $L$ \cite{PhysRevB.80.184402-TDDC,Robin_DQPT_2014}
\begin{equation}
D_{L}(t)=\frac{1}{\chi}\sum_{t^{\prime}=0}^{t-1}\langle j_{\text{tot}}(t^{\prime})j_{\text{tot}}(0)\rangle_{\mu}.
\end{equation}
Here $j_{\text{tot}}(t)$ denotes the extensive spin current \cite{10.21468/SciPostPhys.11.3.051_model}, as defined in the EM. $\chi$ denotes static spin susceptibility, $\chi(\mu)=1+\mu^{-2}-\coth^{2}\mu$. 
We denote by $\langle \bullet \rangle_\mu$ the average in an infinite temperature grand canonical ensemble, with the initial state 
$\rho_\mu^{\rm tot}$,
\begin{equation}\label{eq-gc}
\rho_{\mu}^{\text{tot}}(\boldsymbol{S}_{1},\cdots,\boldsymbol{S}_{L})=\prod_{x=0}^{L-1}\rho_{\mu}(\boldsymbol{S}_{x}),
\end{equation}
where
$ \rho_{\mu}(\boldsymbol{S})=\frac{1}{4\pi}\frac{\kappa}{\sinh \kappa}e^{\kappa S}
$, $\coth \kappa-1/\kappa=\mu$ \cite{krajnik2020kardar-LL}. 
Throughout the paper, we consider vanishing magnetization $\mu = 0$, unless mentioned otherwise.

{\underline{\it Full Counting Stattistics:}\ } 
Additionally, we study 
the cumulative, i.e., time-integrated, spin current density passing through the origin
\begin{equation}
    J(t)=\int_{0}^{t}dt^{\prime}j(0,t^{\prime}) \ .
\end{equation}
The cumulative current $J(t)$ represents the net magnetization transferred between the two halves of a large system that can be
regarded as a fluctuating macroscopic dynamical variable. 
To study the statistical properties of $J(t)$, we introduce the time-dependent probability distribution $\mathcal{P}(J|t)$, which characterizes the statistics of $J(t)$ at time $t$ for an ensemble of initial conditions sampled from the ensemble \eqref{eq-gc}.
More specifically, we analyze ${\cal P}(J | t)$ through its cumulants, $c_{n}(t)=(d/d\lambda)^{n}\log G(\lambda|t)|_{\lambda=0}$.  The function $G(\lambda|t)$ represents 
the moment generating function (MGF), defined as $G(\lambda|t)\equiv\langle e^{\lambda J(t)}\rangle=\int dJ{\cal P}(J|t)e^{\lambda J}$, where the $\langle \bullet \rangle$ is computed in the infinite temperature ensemble. 
Often it is more useful to consider the rescaled cumulants defined as \cite{PhysRevLett.128.090604_Model,PhysRevLett.132.017101_Model}
\begin{align}
   \kappa_{n}(t) = t^{-n/2z}c_{n}(t) = t^{-n/2z}(\frac{d}{d \lambda})^{n}\log G(\lambda|t)|_{\lambda=0}
\label{kappa_definition}
\end{align}
which we focus on in this paper.  Note that $z$ denotes the eqilibrium dynamical exponent and $z=2$ for normal diffusion. Typically, the second rescaled cumulants $\kappa_2(t)$ saturates at $t\rightarrow \infty$, and $\kappa_n(t) \rightarrow 0$, for all $n>2$.
In this regular regime, the central limit property follows from the analytic structure of the MGF, implying that typical fluctuations are normally distributed.

\begin{figure}[t]
	\includegraphics[width=1.0\linewidth]{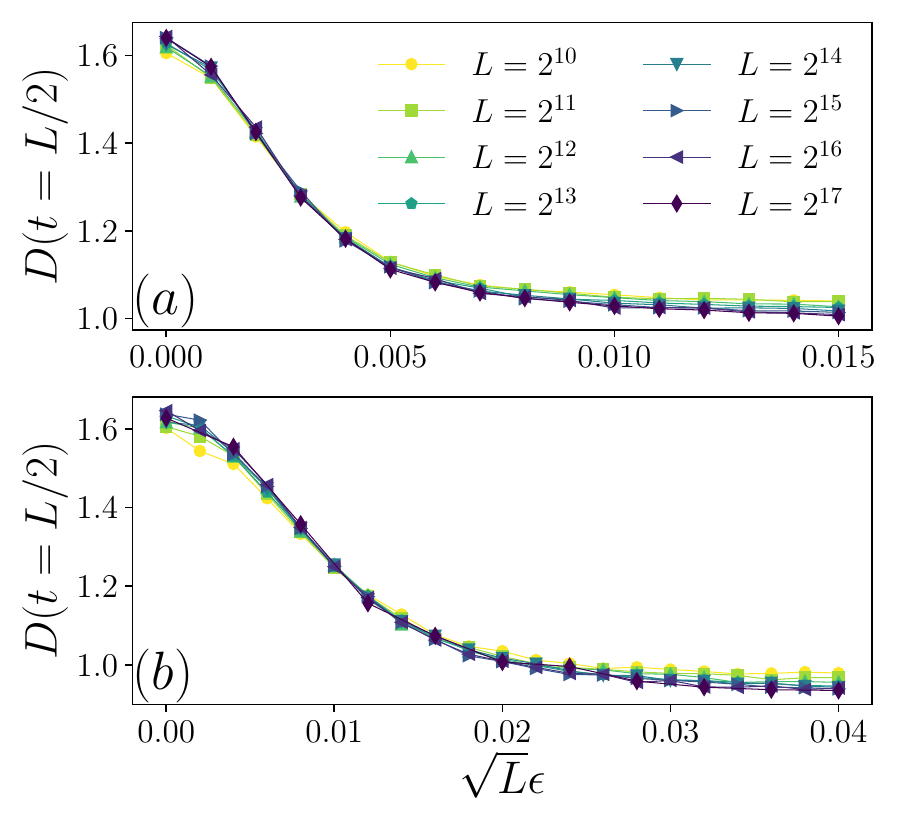}
 \caption{$D(t=L/2)$ versus $\sqrt{L} \epsilon$ for different $L$ in (a) model $A$ and (b) model $B$.}\label{Fig3}
\end{figure}

\begin{figure}[htbp]
\includegraphics[width=1.0\linewidth]{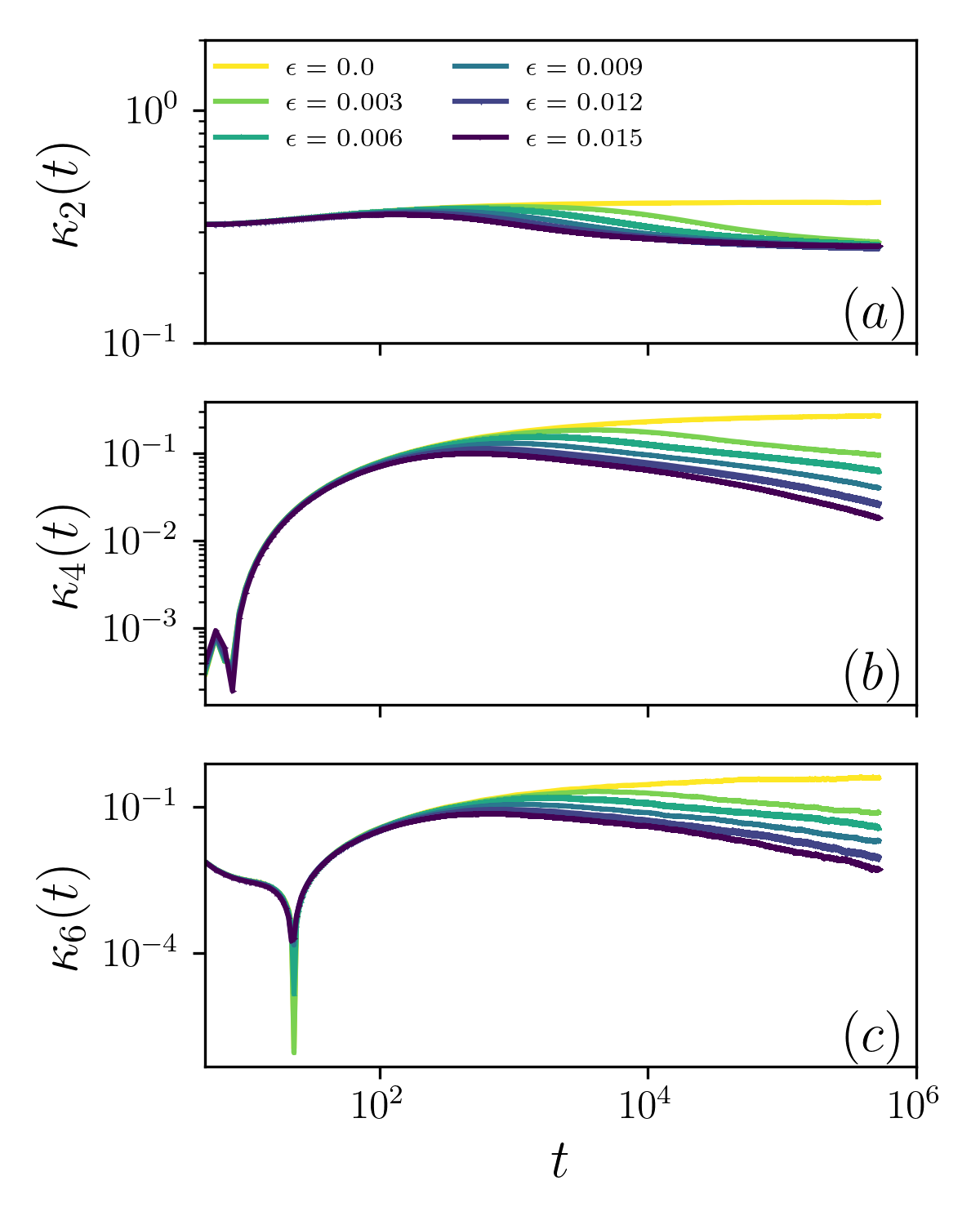}
 \caption{
 Rescaled cumulants of the cumulative current, $\kappa_n(t)$, for different values of $\epsilon$ in model A for system size $L=2^{20}$.}\label{Fig4}
\end{figure}

{\it Numerical results.}
We now present numerical results illustrating the dependence of $D$ and $\kappa_n(t)$ on the integrability-breaking perturbation strength $\epsilon$.

{\underline {\it Results on Diffusion Constant:}\ } 
Figure~\ref{Fig1} shows the time-dependent diffusion constant $D_L(t)$ for various system sizes $L$ and perturbation strengths $\epsilon$ of model A. Results for model B are provided in the End Matter (EM). For the integrable case $\epsilon = 0$, $D_L(t)$ exhibits a rapid initial growth and quickly saturates to a value at a characteristic time scale, both nearly independent of $L$.
At longer times $t \sim L/2$, finite-size effects emerge, leading to a linear increase of $D_L(t)$ \cite{mns1-l19c,PhysRevB.80.184402-TDDC}. Upon applying a perturbation, the integrability is broken and the system enters a chaotic regime, numerical evidence of which in terms of maximum Lyapunov exponent is presented in the EM.
For very tiny strength of integrability-breaking perturbation, $\epsilon = 0.005$, the short-time dynamics remains almost indistinguishable from the integrable case.
However, at later times, $D_L(t)$ begins to decrease, indicating the emergence of a much longer relaxation time scale of the current that governs the eventual saturation.

To examine the scaling of $D_L(t)$ with respect to $\epsilon$, Fig.~\ref{Fig2} presents $D_L(t)$ as a function of $\epsilon^2 t$.
Remarkably, the data collapse onto a single curve up to a crossover time that depends on $\epsilon$, revealing the scaling form
\begin{equation}\label{eq-Dt-f}
D(t)=f(\epsilon^{2}t),\qquad0\ll t\lesssim L/2,
\end{equation}
where $f(\cdot)$ is the scaling function and it is clear from Fig.~\ref{Fig1} that $f(\cdot)$ is independent of $L$.

Since finite-size effects typically emerge around $t \gtrsim L/2$, we define the diffusion constant as
\begin{equation}
    D_L = D(t = L/2),
\end{equation}
which after inserting in Eq.~\eqref{eq-Dt-f} yields
\begin{equation}\label{eq-DL}
    D_L=f(\frac{\epsilon^{2}L}{2})=g(\epsilon\sqrt{L}) .
\end{equation}
To verify this, Fig.~\ref{Fig3} shows $D_L(\epsilon)$ plotted against $\sqrt{L}\epsilon$, where a clear data collapse is observed. Furthermore, this implies
\begin{equation}
\partial_{\epsilon}D_{L}(\epsilon)|_{\epsilon=0}\propto\sqrt{L}\xrightarrow{L\rightarrow\infty}\infty. 
\end{equation}
The above equation shows a nonanalytic dependence of $D_L$ on $\epsilon$ near $\epsilon = 0$ and leads to the conclusion that in the thermodynamic limit, the diffusion constant changes discontinuously upon breaking integrability. Similar phenomenon has recently been reported in quantum spin-1/2 XXZ chain in \cite{DeNArdis_PNAS, Nandy_PRB2022}, suggesting that our findings provide a concrete example of classical-quantum correspondence in the context of transport in near-integrable spin systems. However, it is important also to emphasize that the scenario in quantum case is still under debate \cite{mns1-l19c}, partially due to the limitation of system sizes that are numerically accessible. In contrast, the classical simulations in the present work circumvent these difficulties and demonstrate the phenomenon in a clearer and more convincing way.
Our observations conclusively indicate the emergence of a new diffusion constant in the chaotic regime and is defined as $D_\text{chaos} \equiv \lim_{\epsilon \to 0^{+}} \lim_{L\rightarrow\infty}D_L(\epsilon)$. It differs from the diffusion constant in the integrable case, $D_\text{int} \equiv \lim_{L\rightarrow \infty}D_L(\epsilon = 0)$. However, $D_\text{chaos}$ may depend on the specific form of the integrability-breaking perturbation. 

{\underline {\it Results on Full Counting Statistics:}\ } To further investigate the transport properties, we analyze the rescaled cumulants of the integrated current $\kappa_{n}(t)$ defined in Eq. \eqref{kappa_definition}. For the integrable case ($\epsilon = 0$), as shown in Fig.~\ref{Fig4}, the rescaled cumulants $\kappa_{n}(t)$ tend to saturate to finite values at long times, indicating the emergence of non-Gaussian statistics and thus an anomalous diffusion in this case. This behavior is consistent with the results reported in \cite{PhysRevLett.132.017101_Model}. Additionally, consistency of our results with that in \cite{nonGaussian-XXZ-25} for quantum spin-1/2 XXZ chain once more indicates a possible classical-quantum correspondence\cite{PhysRevB.99.140301-qc,PhysRevB.104.054415-qc,PhysRevB.111.064420-qc,PhysRevResearch.2.013130-qc}. In contrast to integrable case, in the chaotic case ($\epsilon > 0$), $\kappa_{n}(t)$ begins to deviate from the integrable behavior beyond a certain timescale, denoted by $t^*$. While $\kappa_{2}(t)$ remains finite, higher-order cumulants $\kappa_{n>2}(t)$ exhibit a gradual decay at longer times, suggesting that $\kappa_{n>2}(t) \rightarrow 0$. These results imply the restoration of Gaussian statistics under integrability breaking and thus  normal diffusive transport. To gain further insight into the scaling of $t^*$ in the chaotic case, in Fig.~\ref{FigE2} (in the EM) we plot $\tilde{\kappa}_{n}^{\epsilon}(t) \equiv \kappa_{n}^{\epsilon}(t)/\kappa_{n}^{\epsilon=0}(t)$ as a function of the rescaled time $\epsilon^2 t$.  A clear data collapse is observed for all considered cumulants, up to the time scale $t \lesssim t^*$, indicating the scaling $t^* \sim \epsilon^{-2}$.

\underline{\it Conclusion and Outlook:}
We have investigated the magnetization transport at infinite temperature in the space-time
discrete version of integrable classical anisotropic Landau-Lifshitz model, in the presence of (weak) integrability-breaking perturbations. Using large-scale numerical simulations, we computed diffusion constant and analyzed the  higher order cumulants of time-integrated spin current to gain deeper insight into transport characteristics. For the diffusion constant, we established its scaling behavior as a function of system size $L$ and integrability-breaking perturbation strength $\epsilon$. Our results provide a compelling evidence for a discontinuous behavior of diffusion constant near the integrable point as function of $\epsilon$, in the limit $L \rightarrow \infty$. We believe that such a discontinuity could stem from anomalous character of diffusion in integrable systems. Similar conclusion (though challenged in \cite{mns1-l19c}) drawn for quantum spin-1/2 XXZ model \cite{DeNArdis_PNAS, Nandy_PRB2022} indicates a plausible classical-quantum correspondence. Furthermore, our analysis of higher order moments of spin current reveals the emergence of non-Gaussian statistics for the integrable system, indicating that diffusion in integrable system has also anomalous signatures. In contrast, such non-Gaussian behavior is washed out upon the addition of integrability-breaking perturbations (i.e., for finite $\epsilon$), signaling the restoration of normal diffusion in generic systems.

The observed classical-quantum correspondence motivates studying its manifestation in the perturbed regime with strong easy-axis
anisotropy ($\delta \gg 0$ or $\rho \gg 0$
), especially in connection with the 
$z=4$ subdiffusion reported in \cite{DeNArdis_PNAS}. 
Exploring boundary-driven approaches in this setting could clarify their effectiveness and limitations in probing transport via nonequilibrium steady states, as opposed to the dynamical method used here, in line with \cite {Kempa_2025_arXiv}.

\underline{\it Acknowledgment:}
This work has been funded by the Deutsche
Forschungsgemeinschaft (DFG), under Grant No. 531128043, as well as under Grant
No.\ 397107022, No.\ 397067869, and No.\ 397082825 within the DFG Research
Unit FOR 2692, under Grant No.\ 355031190. 
TP acknowledges support by European Research Council (ERC) through the Advanced grant QUEST (Grant Agreement No. 101096208) and from the Slovenian Research and Innovation agency (ARIS) through the program P1-0402.
Additionally, we greatly acknowledge computing time on the HPC3 at the University of Osnabr\"{u}ck, granted by the DFG, under Grant No. 456666331.

\underline{\it Data availability:}\ Research data associated with this
article are openly available \cite{data}.



\bibliographystyle{apsrev4-2-titles}
\bibliography{Ref.bib}

\clearpage
\newpage

 

\section*{End Matter}

\begin{figure}[t]
	\includegraphics[width=1.0\linewidth]{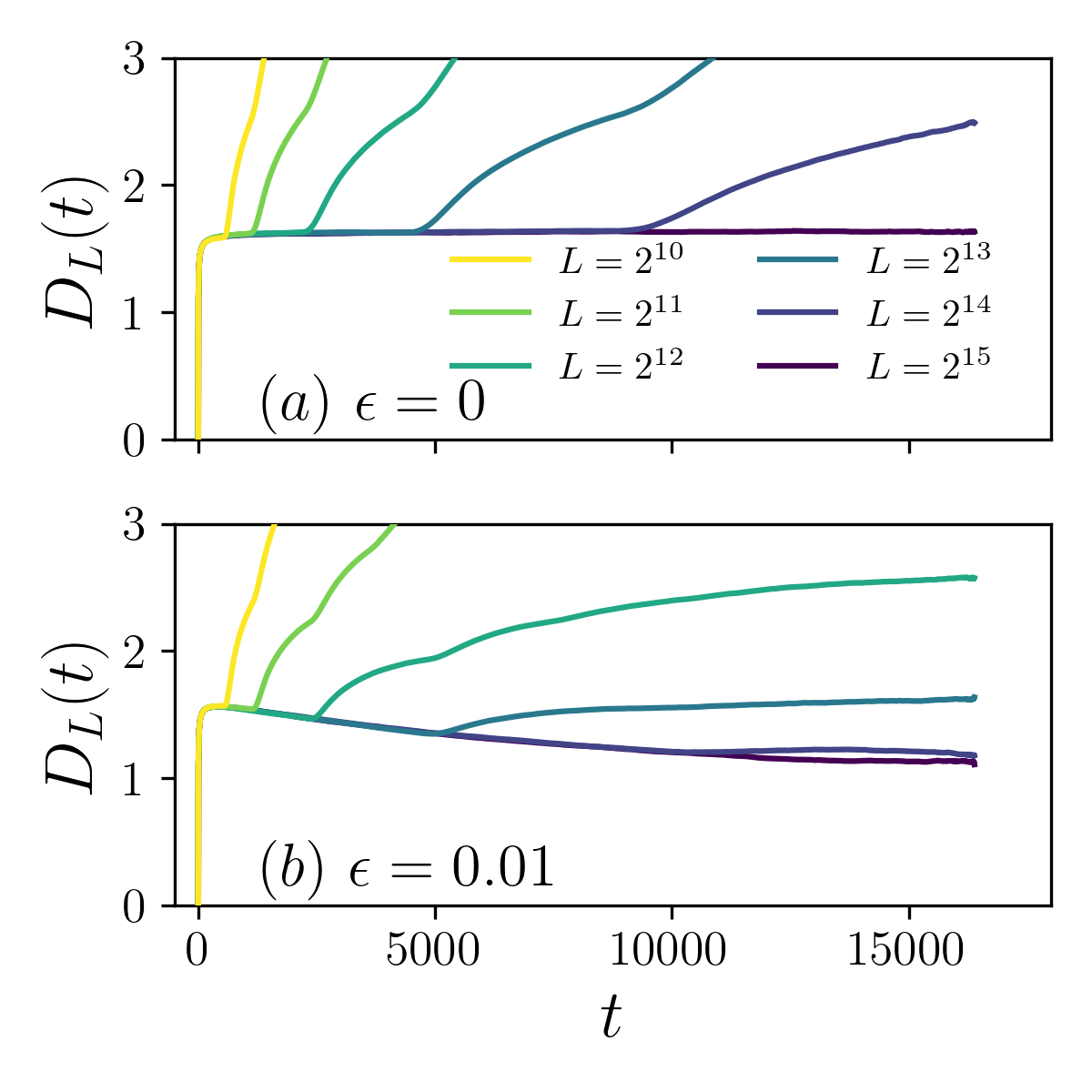}
 \caption{Time-dependent diffusion constant $D_L(t)$ versus $t$ in model $B$ for different system size $L$ for (a) $\epsilon = 0$ and (b) $\epsilon = 0.01$.}\label{FigE1}
\end{figure}

\begin{figure}[t]
\includegraphics[width=1.0\linewidth]{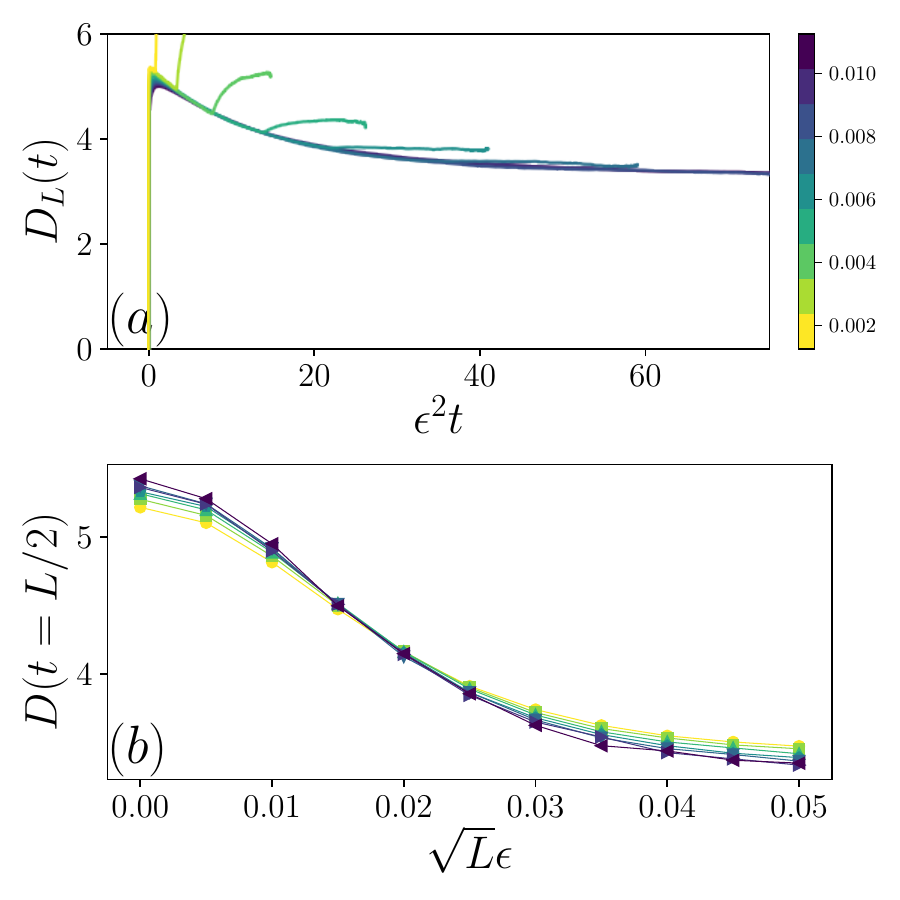}
 \caption{Numerical results in model $A$ for $\rho = 0.5$ and $\tau = 1$.
 (a) $D_L(t)$ versus $\epsilon^2 t$ for $L = 2^{16},\ \rho = 0.5, \tau = 1$. The perturbation strength $\epsilon \in [0, 0.00125]$, with color ranging from yellow to blue as $\epsilon$ increases.
 (b) $D(t=L/2)$ versus $\sqrt{L} \epsilon$ for different $L$.
 }\label{FigE2}
\end{figure}

\begin{figure}[htbp]
\includegraphics[width=1.0\linewidth]{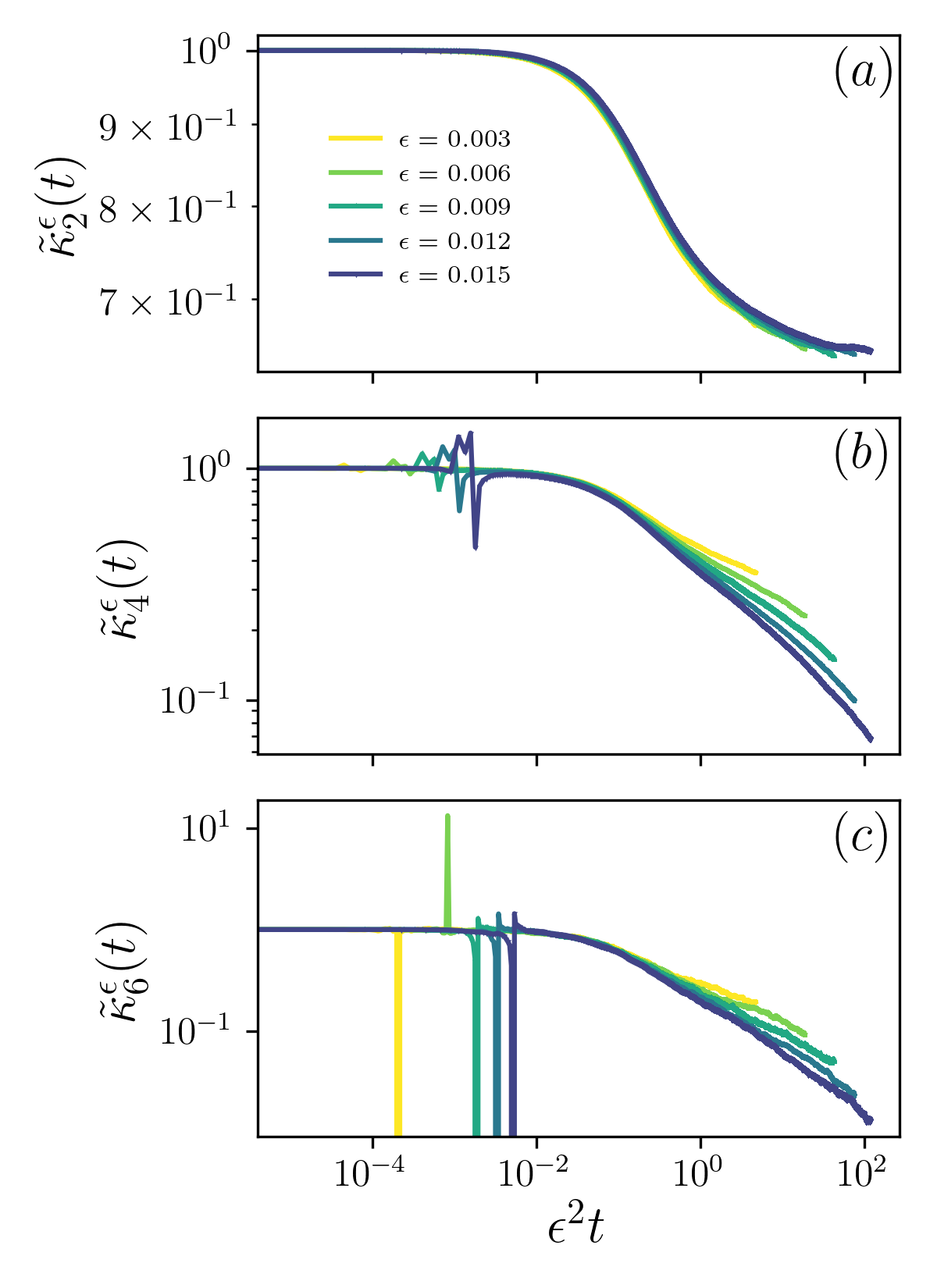}
 \caption{ $
\widetilde{\kappa}^\epsilon_n(t) \equiv 
 \kappa^\epsilon_{n}(t)/\kappa_{n}^{\epsilon=0}(t)$ versus $\epsilon^2 t$ for different $\epsilon$.
}\label{FigE3}
\end{figure}

\subsection*{Definition of the spin current}
To introduce extensive spin current, it is helpful to start with the local discrete space-time N\" other current, which is defined as
\begin{equation}
    j_{x}^{t}\equiv j(x,t)=\frac{1}{\tau}(q(x,t+1)-q(x,t)),
\end{equation}
where
\begin{equation}
q(x,t)\equiv S^{3}(x,t).
\end{equation} 
The extensive spin current $j_{\text{tot}}(t)$ at times $t$ is given by
\begin{equation}
    j_{\text{tot}}(t)=\frac{1}{\sqrt{L}}\sum_{x=0}^{\frac{L}{2}-1}\left(j(2x,t)+j(2x+1,t+\frac{1}{2})\right),
\end{equation}
where $t+\frac{1}{2}$ correspond to observables obtained after applying just a half-time step map
$\varPsi^{\rm odd}_\tau$ after time $t$.

\subsection*{Additional numerical results on diffusion constant}

As discussed in the main text, we present here in Fig. \ref{FigE1}, the time-dependent diffusion constant $D_{L}(t)$ for various system sizes $L$ and perturbation strengths $\epsilon$ for model B. This figure, both in its contents and conclusions, is analogous to Fig. \ref{Fig1}. As in Fig. \ref{Fig1} in the main text, here also we see that i) for the integrable case ($\epsilon=0$), $D_L(t)$ saturates rapidly to a value independent of $L$ and finite size effects creep in roughly at $t \sim L/2$. ii) For integrability-broken case ($\epsilon=0.01$), as before, the short-time dynamics remains more or less indistinguishable from the integrable case whereas $D_{L}(t)$ at later times begins to decrease, indicating
the emergence of a much longer relaxation scale that governs the eventual saturation. 

To further examine the generality of Eqs.~\eqref{eq-Dt-f} and \eqref{eq-DL}, we consider, in addition to the case $\rho = 1.0$ discussed in the main text, an alternative parameter value $\rho = 0.5$ for model $A$. The results, shown in Fig.~\ref{FigE2}, are similar to those in Figs.~\ref{Fig2} and \ref{Fig3}, supporting the validity of the scaling in Eqs.~\eqref{eq-Dt-f} and \eqref{eq-DL}.

\begin{figure}[htbp]
\includegraphics[width=1.0\linewidth]{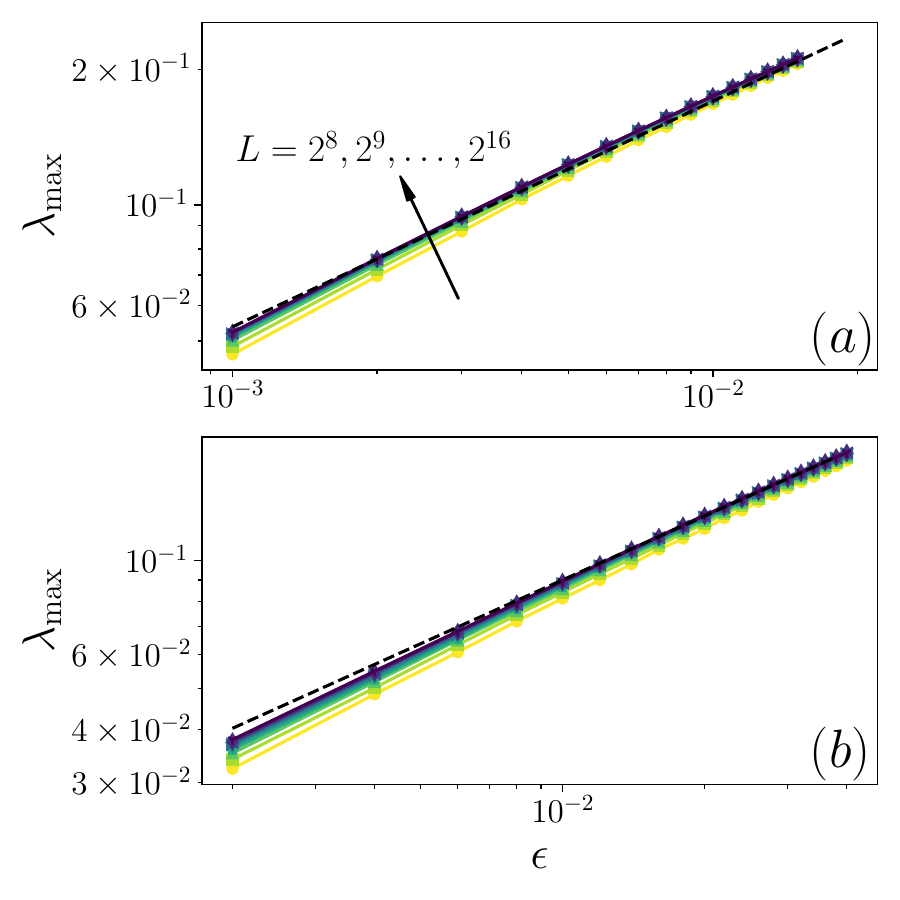}
 \caption{Maximum Lyapunov exponent $\lambda_{\text{max}}$ versus $\epsilon$ for different $\epsilon$ in (a) model A and (b) model B. Dashed lines indicate the scaling $\propto \sqrt{\epsilon}$}\label{FigS3}
\end{figure}
\subsection*{Additional numerical results on higher order cumulant of time-integrated current distribution}

As discussed in the main text, the rescaled cumulant $\kappa_{n}(t)=t^{-n/2z}c_{n}(t)$ begins to deviate from the integrable behavior ($\epsilon=0$) in the presence of integrability-breaking i.e., $\epsilon \ne 0$ beyond a time scale $t^{*}$. To investigate how how this $t^{*}$ depends upon $\epsilon$, we plot $\tilde{\kappa}_{n}^{\epsilon}(t) \equiv \kappa_{n}^{\epsilon}(t)/\kappa_{n}^{\epsilon=0}(t)$ in Fig. \ref{FigE2}. A clear data collapse up to time scale $t \lesssim t^*$, indicates the scaling behavior as $t^* \sim \epsilon^{-2}$.

\subsection*{Maximum Lyapunov exponents}
To investigate the integrability-to-chaos transition, we calculate the maximum Lyapunov exponent $\lambda_{\text{max}}$ as a function of the perturbation strength $\epsilon$.
In both models, we can see that, $\lambda_{\text{max}}$ tends to converge with increasing system size $L$ for any fixed $\epsilon$.
Furthermore, a scaling law $\lambda_{\text{max}} \propto \sqrt{\epsilon}$ is observed for all considered $L$, which is more pronounced for $\epsilon$ not extremely small.

\end{document}